\documentclass[graybox]{svmult}

\pdfoutput=1
\usepackage{mathptmx}       
\usepackage{helvet}         
\usepackage{courier}        
\usepackage{type1cm}        
\usepackage{makeidx}         
\usepackage{graphicx}        
\usepackage{multicol}        
\usepackage[bottom]{footmisc}

\makeindex             

\begin{document}

\title*{Evidence for Pulsation-Driven Mass Loss from $\delta$~Cephei}
\author{M. Marengo, N. R. Evans, L. D. Matthews, G. Bono, P. Barmby,
  D. L. Welch, M. Romaniello, K. Y. L. Su, G. G. Fazio \& D. Huelsman}
\authorrunning{Marengo et al.}
\institute{M. Marengo \at Iowa State University, Dept. of
  Physics, Ames, IA 50011, \email{mmarengo@iastate.edu}}
%
%
\maketitle


\abstract{We found the first direct evidence that the Cepheid class
  namesake, $\delta$~Cephei, is currently losing mass. These
  observations are based on data obtained with the \emph{Spitzer}
  Space Telescope in the infrared, and with the Very Large
  Array in the radio. We found that $\delta$~Cephei is associated with 
  a vast circumstellar structure, reminiscent of a bow shock. This
  structure is created as the wind from the star interacts with the
  local interstellar medium. We measure an outflow velocity of
  $\approx 35.5$~km\,s$^{-1}$ and a mass loss rate of $\approx 10^{-7}$ --
  $10^{-6} \ M_\odot$~yr$^{-1}$. The very low dust content of the outflow
  suggests that the wind is possibly pulsation-driven, rather than
  dust-driven as common for other classes of evolved stars.}

\abstract*{We found the first direct evidence that the Cepheid class
  namesake, $\delta$~Cephei, is currently losing mass. These
  observations are based on data obtained with the \emph{Spitzer}
  Space Telescope in the infrared, and with the Very Large
  Array in the radio. We found that $\delta$~Cephei is associated with 
  a vast circumstellar structure, reminiscent of a bow shock. This
  structure is created as the wind from the star interacts with the
  local interstellar medium. We measure an outflow velocity of
  $\approx 35.5$~km\,s$^{-1}$ and a mass loss rate of $\approx 10^{-7}$ --
  $10^{-6} \ M_\odot$~yr$^{-1}$. The very low dust content of the outflow
  suggests that the wind is possibly pulsation-driven, rather than
  dust-driven as common for other classes of evolved stars.}

\section{Implications of mass-loss processes in the Cepheid phase}
\label{sec:1}

Cepheids hold the key to the cosmological distance scale. Thanks
to the period--luminosity relation (Leavitt law, \cite{leavitt1908}),
they are the first rung in the ladder we use to measure the size and
age of the universe. They are also a benchmark for intermediate-mass
stellar evolution models. Despite their importance, however, there are
still outstanding puzzles in their theoretical understanding. In particular,
the mass predicted by evolutionary models is significantly larger than
the mass estimated by pulsation theory, or directly measured in binary
systems \cite{cox1980, caputo2005}.

Recent calculations \cite{neilson2011, bono2006} show that the Cepheid
mass discrepancy can be solved by evolutionary models including
convective core overshoot and mass loss. These models can dramatically
lower the predicted initial mass of a Cepheid, without preventing the
star from crossing the instability strip in the so-called ``blue
loop'' characteristic of the Cepheid phase. The main difficulty with
these models is that both overshooting and mass loss rate need to be
included as free parameters, rather than independently derived from
stellar physics by first principles. 

\begin{figure}[t]
\includegraphics[width=0.99\textwidth]{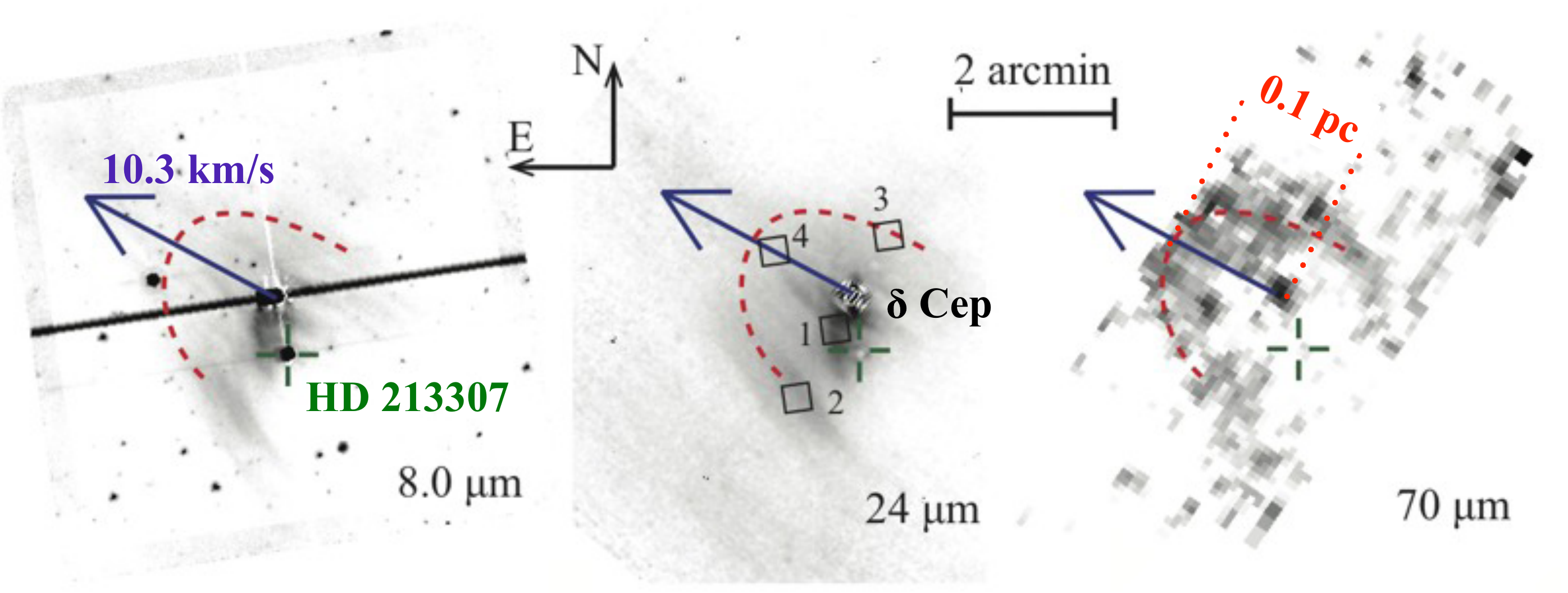}
\caption{\emph{Spitzer} images of the $\delta$~Cephei nebula. The
  dashed line outlines the bright 70~$\mu$m arc. The arrow indicates
  the direction of the space velocity relative to the local ISM. The
  boxes in the middle panel show the locations where we probed the
  extended emission flux ratios.}
\label{fig:1}
\end{figure}

Our limited understanding of stellar convection is unlikely to improve
in the short term, but mass loss can be directly probed with
observations. Understanding Cepheid mass loss is important not only
because it affects their evolution, but also for its effect on the
distance scale. Standard candles can be effectively used only if their
apparent luminosity can be precisely measured. Mass loss, which
surrounds the star with circumstellar material, is a source of both
visible extinction and infrared excess. Both factors introduce noise
in the Cepheid period--luminosity relationship, and should be taken into
account if we want to reach our goal of 2 percent systematic accuracy in the
determination of the Hubble constant \cite{freedman2011}.

While some Cepheids are known to be associated with
reflection nebul\ae{} (e.g. RS Pup), it has been difficult to prove
that the circumstellar material is the result of a stellar
wind.  Intermediate-mass stars have a relatively short life.  
Such Cepheids tend to be found in regions where they are formed and where
the interstellar medium (ISM) is dense.  It is therefore possible that 
the excess emission measured in the infrared \cite{mcalary1986, neilson2009, 
barmby2010} and in the UV \cite{deasy1988} may originate in the nearby
ISM.  Recent high angular resolution observations (\cite{gallenne2011,
 kervella2009, merand2007} and references therein) have revealed the
presence of compact circumstellar shells around several Cepheids. The
proximity of these shells to the pulsating stellar photosphere (as
close as a 2--3 stellar radii) may indicate the existence of some mass
loss mechanism triggered by shocks associated with stellar
pulsation. Direct evidence of a large-scale mass loss process capable
of explaining the Cepheid mass discrepancy has, however, eluded all
observational efforts.

Our \emph{Spitzer} observations of the Cepheid namesake $\delta$~Cephei 
\cite{marengo2010} may, however, have just filled this gap. We
found a large-scale structure which is best explained as an infrared
bow shock resulting from the interaction of a strong stellar wind
with the local ISM. Our VLA 21-cm line data \cite{matthews2011}
confirm this hypothesis and provide the first measurement of a
Cepheid wind velocity as well as strong constraints on its current 
mass-loss rate.

\section{The $\delta$~Cephei nebula}
\label{sec:2}

\begin{figure}[t]
\includegraphics[width=0.99\textwidth]{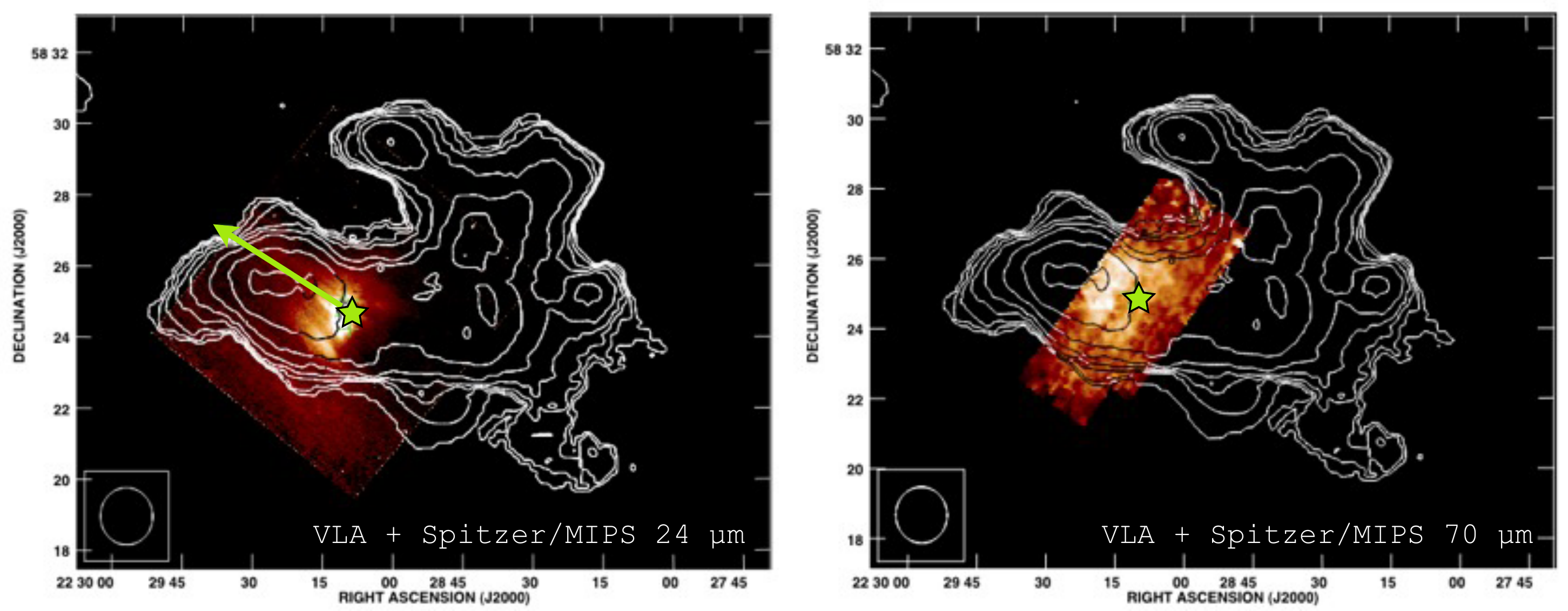}
\caption{VLA H\textsc{i} 21-cm line map of the $\delta$~Cephei
  nebula overlayed on our 24 and 70~$\mu$m \emph{Spitzer} images. The
  location of the star is marked by the green symbol, and the
  $\delta$~Cephei space velocity is indicated by the green arrow.}
\label{fig:2}
\end{figure}

Fig.~\ref{fig:1} shows our \emph{Spitzer} images of the
$\delta$~Cephei nebula. The emission at 8.0 and 24~$\mu$m is stronger
between the star and its widely separated companion (HD~213307, a B7-8
main sequence star), and is enclosed within an arched structure
detected at 70~$\mu$m. This structure is aligned with the space
velocity of the star ($\approx 10.3$~km\,s$^{-1}$ and P.A. $\approx
58.3^\circ$). Flux ratios measured in different positions along the
arc reveal a temperature of about 100~K. The composition, which
has low PAH content with respect to the average ISM, is consistent with 
this structure being formed by a strong stellar wind pushing against the
local ISM, as is common in stellar bow shocks.

This interpretation is confirmed by our VLA H\textsc{i} 21-cm line
mapping of the region (Fig.~\ref{fig:2}). The data reveal a large
($\sim 1$~pc) nebula with a head-tail morphology, consistent with
circumstellar ejecta shaped by the interaction between a stellar wind
and the ISM. The bulk of the emission overlaps with the arc structure
detected in the infrared, with a trailing tail in the opposite
direction to the stellar space velocity. By fitting the H\textsc{i}
line in the velocity beams not contaminated by background emission, we
derived an outflow velocity of $\simeq 35.5$~km\,s$^{-1}$. This is the first
actual measurement of the wind velocity for a Cepheid. It is worth
noting that this velocity is significantly smaller than the expected
escape velocity from the star ($\sim 200$~km\,s$^{-1}$). Based on this
measurement, the dynamical age of the structure is $\sim 10^5$~yr,
consistent with the expected duration of the Cepheid phase in this
star.

\section{Evidence for pulsation-driven mass loss} 
\label{sec:3}

The VLA data, combined with our infrared observations, provide strong
constraints on the current mass-loss rate of the star. The total flux
density of the H\textsc{i} nebula, and detailed fitting of the
21-cm line, are consistent with a $\delta$~Cephei mass-loss rate of
$\approx 10^{-6} \ M_\odot$ yr$^{-1}$. This value may be considered an upper
limit, since a fraction of the H\textsc{i} may have been swept out
from the ISM, rather than having been lost from the star by the wind.
An alternative estimate can be derived from the observed
stand-off distance of the bow shock-like structure, together with ram
pressure balance arguments and our measured wind velocity. This
provides a lower limit of $\approx 10^{-7} \ M_\odot$ yr$^{-1}$, by adopting
a conservative estimate of the local ISM density. Models show that a
mass-loss rate in this range is sufficient to solve the Cepheid mass
discrepancy if sustained during the time the star is crossing the
instability strip.

The total flux density detected with the VLA corresponds to a total
H\textsc{i} mass of $\approx 0.07 \ M_\odot$. Comparison with the dust
mass derived from the infrared flux detected by \emph{Spitzer} implies
a gas-to-dust mass ratio of $\approx 2300$. This value is
significantly higher than the canonical ratio observed in 
dust-driven winds of evolved giant stars and in the ISM. This is
consistent with the low PAH content measured for the $\delta$~Cephei
nebula, and is further evidence that the mass-loss process acting in
this star is not dust-driven.

Additional evidence is provided by the observation that the measured wind
velocity is much smaller than the escape velocity. This condition
requires the existence of some regulatory process capable of lifting
the stellar atmosphere, rather than accelerating the flow. It also
requires that most of the energy added to the wind must be in the form
of momentum, rather than heat. In evolved cool stars this momentum
transfer is provided by the friction of dust grains accelerated by
radiation pressure. In the $\delta$~Cephei wind there is not enough
dust to support this process and, due to the higher effective
temperature of the star, the dust cannot form close enough to the
stellar photosphere to effectively trigger this process. An
alternative source for the required mechanical energy could, however, be
provided by the pulsation of the star, a known source of strong shocks
periodically crossing the stellar atmosphere and chromosphere
\cite{marengo2002}, suggesting that this could be a pulsation-driven
wind.

\end{document}